\documentclass{aa}  

\usepackage{graphicx}
\usepackage{txfonts}

\usepackage{natbib}
\bibpunct{(}{)}{;}{a}{}{,} 

\usepackage[colorlinks=true, urlcolor=blue, colorlinks=true, citecolor=blue, linkcolor=blue]{hyperref}

\begin{document} 

   \title{Past activity of Sgr A$^\star$ is unlikely to affect the local cosmic-ray spectrum up to the TeV regime}
   
   \author{M. Fournier
          \inst{1,}\inst{2},
          J. Fensch\inst{1} and B. Commerçon\inst{1} 
          }

   \institute{Centre de Recherche Astrophysique de Lyon UMR5574, ENS de Lyon, Univ. Lyon1, CNRS, Université de Lyon, 69007 Lyon, France \\ \email{jeremy.fensch@ens-lyon.fr; benoit.commercon@ens-lyon.fr} \and Universität Hamburg, Hamburger Sternwarte, Gojenbergsweg 112, 21029 Hamburg, Germany \\
              \email{martin.fournier@hs-uni.hamburg.de}
             }

   \date{\today}

  \abstract
   {The presence of kiloparsec-sized bubble structures in both sides of the Galactic plan suggests active phases of Sgr A$^\star$, the central supermassive black hole of the Milky--Way in the last 1--6 Myr. The contribution of such event on the cosmic-ray flux measured in the solar neighborhood is investigated with numerical simulations.}
   {We evaluate whether the population of high--energy charged particles emitted by the Galactic Center could be sufficient to significantly impact the CR flux measured in the solar neighborhood.}
   {We present a set of 3D magnetohydrodynamical simulations, following the anisotropic propagation of CR in a Milky -- Way like Galaxy. Independent populations of cosmic-ray are followed through time, originating from two different sources types, namely Supernovae and the Galactic Center. To assess the evolution of the CR flux spectrum properties, we split these populations into two independent energy groups of 100 GeV and 10 TeV.}
   {We find that the anisotropic nature of cosmic-ray diffusion dramatically affects the amount of cosmic-ray energy received in the solar neighborhood. Typical timescale to observe measurable changes in the CR spectrum slope is of the order 10 Myr, largely surpassing estimated ages of the Fermi bubbles in the AGN jet-driven scenario.}
   {We conclude that a cosmic-ray outburst from the Galactic center in the last few Myr is unlikely to produce any observable feature in the local CR spectrum in the TeV regime within times consistent with current estimates of the Fermi bubbles ages.}


   \maketitle
%

\section{Introduction}

Building a consistent description of the cosmic-ray (CR) spectrum has been a challenging theoretical task for decades. While it is well established that these relativistic charged particles are accelerated by magnetic shocks \citep{Fermi49,Bell78,Hillas84,Blandford87}, a comprehensive understanding of the relative contributions of all these sources is still lacking. It is thought that the Galactic cosmic-rays (GCR) are dominating the overall CR population for energies below $\sim 10^{18}$ eV, while substructures in the CR spectrum, the so-called \textit{knees} and \textit{ankle}, are considered as indicators for a transition between galactic and extragalactic CR \citep{Aloisio_2012}. 

In common scenarios, GCR find their origin in magnetic shocks produced by Supernovae Remants (SNR) \citep{Strong_2007}. The total thermal energy released by such event, of order $10^{51}$ erg, is believed to be partially converted into CR with a yield of $\sim 10$ \%, and with energies up to a few 10-100 TeV \citep{Lagage83}.

The case for the existence of other significant CR sources within the Galaxy has recently been opened by the discovery of large hemispherical structures above and below the Galactic plane by the ROSAT, Fermi and eRosita space telescopes \citep{Bland_2003,Su_2010}. The soft X-ray component of these so-called \textit{Fermi bubbles} (FB) consists of spherical bright shells extending up to 14 kiloparsec (kpc) in both sides the galactic plane. The edge of these shells seems to coincide with that of a gamma-ray component, filling its inner volume with a rather uniform surface brightness. Two main scenarios are debated to explain these features: starbust-driven disk outflows \citep{Sofue_2000} or a past activity of the central supermassive black hole (SMBH), Sgr A$^\star$ \citep{Guo_2012}. Both these scenarios involve a total energy injection $E_\text{F}$ of at least $10^{55}$ erg, with the latter ranging between $10^{56}$ and $10^{57}$ erg \citep{Guo_2012,Yang_2022}. Also, hydrodynamical simulations allow to put a constraint on their age, which is estimated to be $\sim$ 1 -- 6 Myr for the SMBH jet-driven scenarios \citep{Guo_2012,Zhang_2020}, and up to 50 Myr for nuclear star formation driven bubbles \citep{Miller_2016}. Whether the expanding X-ray shells and the volume-filling gamma-ray component correspond to a single event is still unclear since accretion from SMBH is a stochastic phenomenon which can leads to repetitive emissions of outbursts separated by timescales down to $\sim 10^5$ yr \citep{King_2015}.

Theoretical studies suggest that two main scenarios can be proposed to explain the observations of the FB, where the gamma-ray emission are the result of either (i) CR protons interaction with the surrounding medium (“hadronic” model) or (ii) inverse Compton scattering by CR electrons (“leptonic” model) \citep{Yang_2012}. In both cases, the event leading to the formation of the FB is expected to have accelerated a large quantity of CR. Whether this specific population of proton and electron CR could have a significant contribution to the CR spectrum measured on Earth remains unclear. Theoretical work following the isotropic propagation of CR in a simplified Milky Way (MW) model concludes that only a small fraction of the assumed injected energy $E_\text{F}$ is found to be sufficient to significantly affect the CR spectrum measured in the solar neighborhood \citep{Jaupart_2018}. \\

In this paper, we revisit these conclusions using a realistic magnetohydrodynamical numerical model of the MW. In particular, our model includes magnetic fields and CR diffusion along the magnetic field lines. In details, we propose to study the diffusion of CR accelerated by SNR or by a central source. The paper is organised as follow. In Sect.~2 we present the numerical setup. In Sect.~3 we present how populations of different origin diffuse into the Galaxy for two different energies, and discuss the implications for the CR spectrum observed in the solar neighborhood. Conclusions and perspectives are presented in Sect.~4.

\section{Simulations and method}
\subsection{Magnetohydrodynamics with anisotropic cosmic-ray diffusion}
To simulate the magnetohydrodynamical evolution of an isolated MW Galaxy, we used the adaptative mesh refinement code {\scshape{Ramses}} \citep{Fromang_2006}. We use the Harten–Lax–van Leer Discontinuity (HLLD) \citep{HLLD} Riemann solver to calculate the MHD flux. Computation of the CR energy density flux is performed using the implicit numerical solver introduced in \citet{Dubois_2016} and \citet{Dubois_2019}, which we modified to support several CR groups. The full system is then described by the following set of equations:

\begin{equation}
    \begin{aligned}
        &\frac{\partial \rho}{\partial t} + \nabla \cdot [\rho \mathbf{u}] = 0, \\
        &\frac{\partial \rho \mathbf{u}}{\partial t} + \nabla \cdot \left[ \rho \mathbf{u} \otimes \mathbf{u} + P_{\text{tot}} \mathbb{I} - \frac{\mathbf{B}\otimes\mathbf{B}}{4 \pi} \right] = 0, \\
        &\frac{\partial E_{\text{tot}}}{\partial t} + \nabla \cdot  \left[ \mathbf{u}(E_{\text{tot}} + P_{\text{tot}}) - \frac{(\mathbf{u}\cdot\mathbf{B})\mathbf{B}}{4 \pi} \right] \\ 
        &= -P_{\text{CR}} \nabla \cdot \mathbf{u} + \nabla \cdot (\mathbb{D} \nabla E_{\text{CR}}) - \mathcal{L} (\rho,T), \\
        &\frac{\partial \mathbf{B}}{\partial t} - \nabla \times [\mathbf{u} \times \mathbf{B}] = 0, \\
        &\frac{\partial E_{\text{CR}}}{\partial t} + \nabla \cdot ({\mathbf{u} E_{\text{CR}}}) = -P_{\text{CR}} \nabla \cdot \mathbf{u} + \nabla \cdot (\mathbb{D} \nabla E_{\text{CR}}). \\
    \end{aligned}
\label{MHD}
\end{equation}

Here, $\rho$ is the gas density, $\mathbf{u}$ its velocity field, $\mathbf{B}$ is the magnetic field vector, $P_{\text{tot}}$ is the total pressure including the gas pressure $P=(1-\gamma)E$, the magnetic field pressure $P_{\text{mag}} = B^2/(8\pi)$ and the CR pressure $P_{\text{CR}}=(\gamma_{\text{CR}}-1)E_{\text{CR}}$, where $\gamma$ and $\gamma_{\text{CR}}$ are the adiabatic index of the gas and the CR respectively. The CR being here considered as a relativistic fluid, we take $\gamma_{\text{CR}}=4/3$, while the gas is non-relativistic and thus caracterized by the index $\gamma=5/3$. $E$ is the gas internal energy density, $E_{\text{CR}}$ is the CR energy density and $E_{\text{tot}}$ designates the total energy density $E_{\text{tot}}=E + 0.5 \rho u^2 + B^2/(8\pi) + E_{\text{CR}}$. $\mathcal{L}(\rho,T)$ is a cooling function modelling the heating and cooling processes taking place inside the ISM. $\mathbb{D}$ describes the diffusion tensor of the CR, and can be decomposed following:

\begin{equation}
    \mathbb{D} = D_\perp \delta_{ij} - (D_\perp - D_\parallel) \, \mathbf{b} \times \mathbf{b},
\end{equation}

where $\mathbf{b}=\mathbf{B} \ /\parallel \mathbf{B} \parallel$ is the normalised magnetic field vector and $D_\parallel$ is the parallel diffusion coefficient, accounting for the diffusion along magnetic field lines. $D_{\perp}$ is the perpendicular diffusion coefficient, effectively taken as $10^{-2} \times D_\parallel$, which account for the diffusion of CR perpendicularly to the magnetic field lines.

\subsection{Initial conditions}

The initial conditions are similar to those in \citet{Renaud_2013}. We initially set a box of width $L_{\text{box}}=120$ kpc filled with an axisymmetric distribution of gas and stellar particles. Particles position and velocity distributions are performed using the Multi-Gaussian Expansion (MGE) method \citep{Emsellem_1994}, which uses a set of Gaussian functions to model the various component of the MW based on a given input, here chosen as the Besançon model \citep{Robin_2003}. The magnetic field is initially seeded with a toroidal component of amplitude 1 $\mu$G. 

A spherically symmetric dark matter halo is modelled by the following density profile:

\begin{equation}
    \begin{aligned}
        \rho(r) \propto \left\{ 1 + \left( \frac{r}{r_c} \right)^2 \right\}^{-1},
    \end{aligned}
\end{equation}

with a characteristic radius $r_c$ of 2.7~kpc and up to a truncation radius at 50~kpc, using $4 \times 10^6 \text{ particles of } 1.13 \times 10^5~\text{M}_\odot$. A pre-existing star population of $4.6 \times 10^{10}~\text{M}_\odot$ is modelled with $4 \times 10^6 \text{ particles of } 1.15 \times 10^4~\text{M}_\odot$. This population is decomposed into a bulge, a spheroid, a thick and a thin disk following \citet{Renaud_2013}. In addition, a particle of mass $4 \times 10^{6}~\text{M}_\odot$ is located a the center of the box to model the SMBH. The gas component is modeled by an exponential profile with radial (resp. vertical) scalelength of 10~kpc (0.25~kpc) and truncation at 30~kpc (2.5~kpc), with a total mass of $5.94 \times 10^9~\text{M}_\odot$.

The coarsest cells have a size of 937.5~pc. We allow refinement down to 14~pc following two criteria. We refine a cell if (i) there are more than 20 particles from initial conditions or more than $2 \times 10^4~\text{M}_\odot$ total mass in the cell, or (ii) the local thermal Jeans length is smaller than four times the cell size \citep{Truelove_1997}.

As gas and stars do not initially share the same gravitational potential, and we first need to simulate a phase of “warm relaxation” for about 1 Gyr at the beginning of our simulation with no star formation and isothermal equation of state at $T = 5 \times 10^4$~K so that any structure formed during the relaxation of the gas+particles system can dissipate. During this phase, instabilities lead to the formation of the asymmetric structures of the Galaxy, namely the bar and the spiral arms. 

Once these structures are formed, we activated gas cooling and heating and star formation. Gas cooling is computed via cooling tables tabulated at solar metallicity available in the public version of {\scshape{Ramses}}, with a temperature floor of 50~K. Star formation is modelled by a density based criterion: each cell with gas density above 1 cm$^{-3}$ and colder than $2 \times 10^4$~K forms stars following the \citet{Schmidt1959} law with 10\% efficiency per free-fall time. New stars have a mass of $10^4~\text{M}_\odot$.

\begin{figure}[h!]
    \centering
    \includegraphics[width=0.5\textwidth]{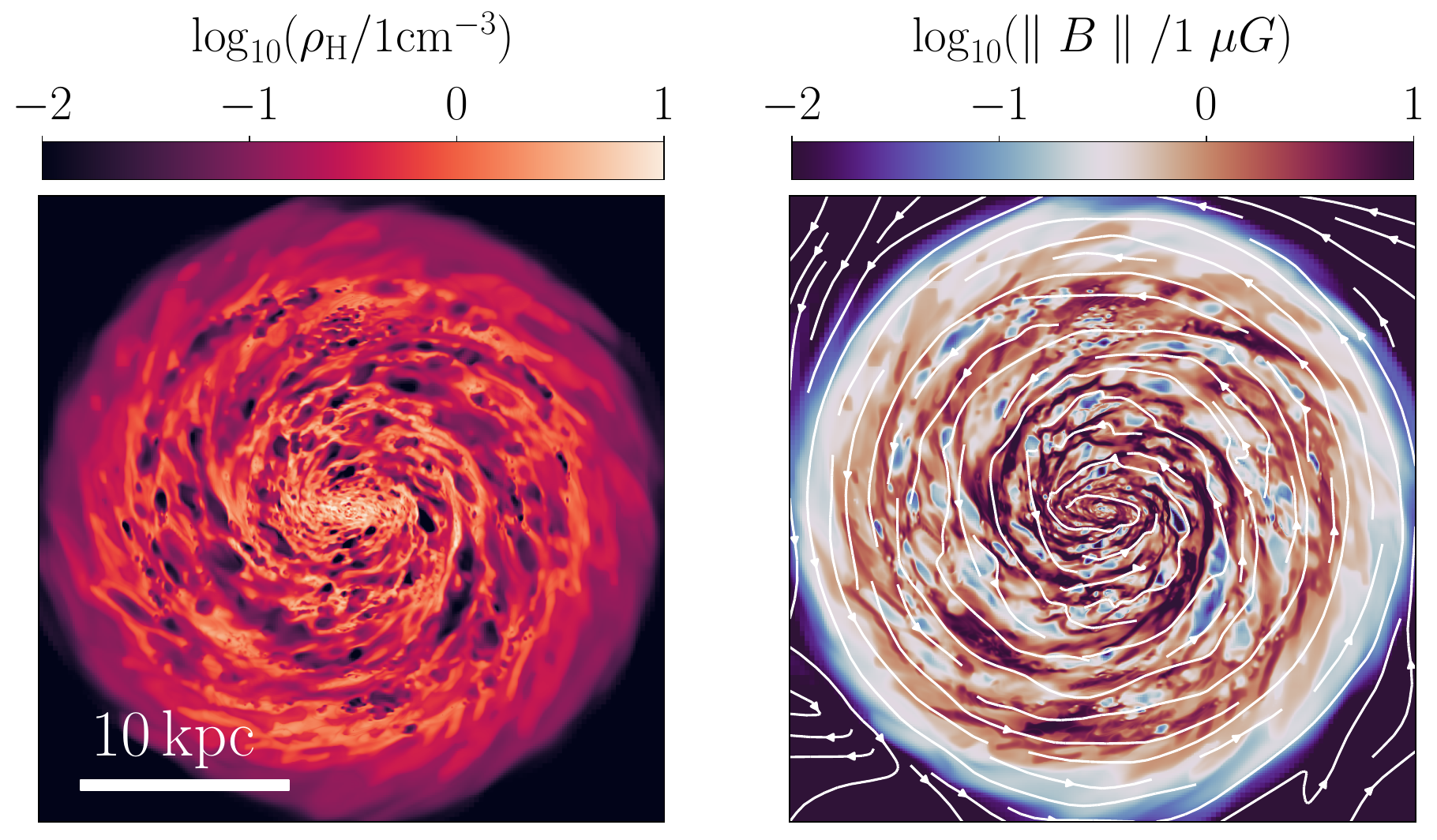}
    \caption{Initial conditions for CR injection. The left panel shows the hydrogen density, while the right panel shows the magnetic field amplitude overlaid with its associated field lines.}
    \label{fig:condini}
\end{figure}

After an additional $\sim 50$ Myr, the star formation rate (SFR) reaches a realistic order of magnitude (i.e. around $1 \, \text{M}_\odot / \text{yr}$). The corresponding gas density and magnetic fields maps are shown in Fig. \ref{fig:condini}. CR injection from SNe is then turned on and the simulation is continued for an extra 5 Myr so that the CR injected from the SNe have enough time to diffuse significantly. Time at the end of this phase will as of now be referred as $t=0$.

\subsection{Stellar feedback}
\label{feedback}

We include feedback from type II SNe. Injection is performed for stellar particles with an age of more than 10 Myr and a mass superior to 10 $\text{M}_\odot$. At each injection event, mass and specific thermal energy are injected in the parent gas cell of the stellar particle following:

\begin{equation}
    \label{stellarfeedback}
    \begin{aligned}
        &m_{\rm{SN}} = f_{\rm{SN}} \times m_\star, \\
        &e_{\text{SN}} = 10^{51} \, \text{erg} / (10 \, \text{M}_\odot),
    \end{aligned}
\end{equation}

\noindent where $m_\star$ is the mass of the stellar particle, and $f_{\text{SN}} = 0.2$.

\begin{figure*}[h!]
    \centering
    \includegraphics[width=\textwidth]{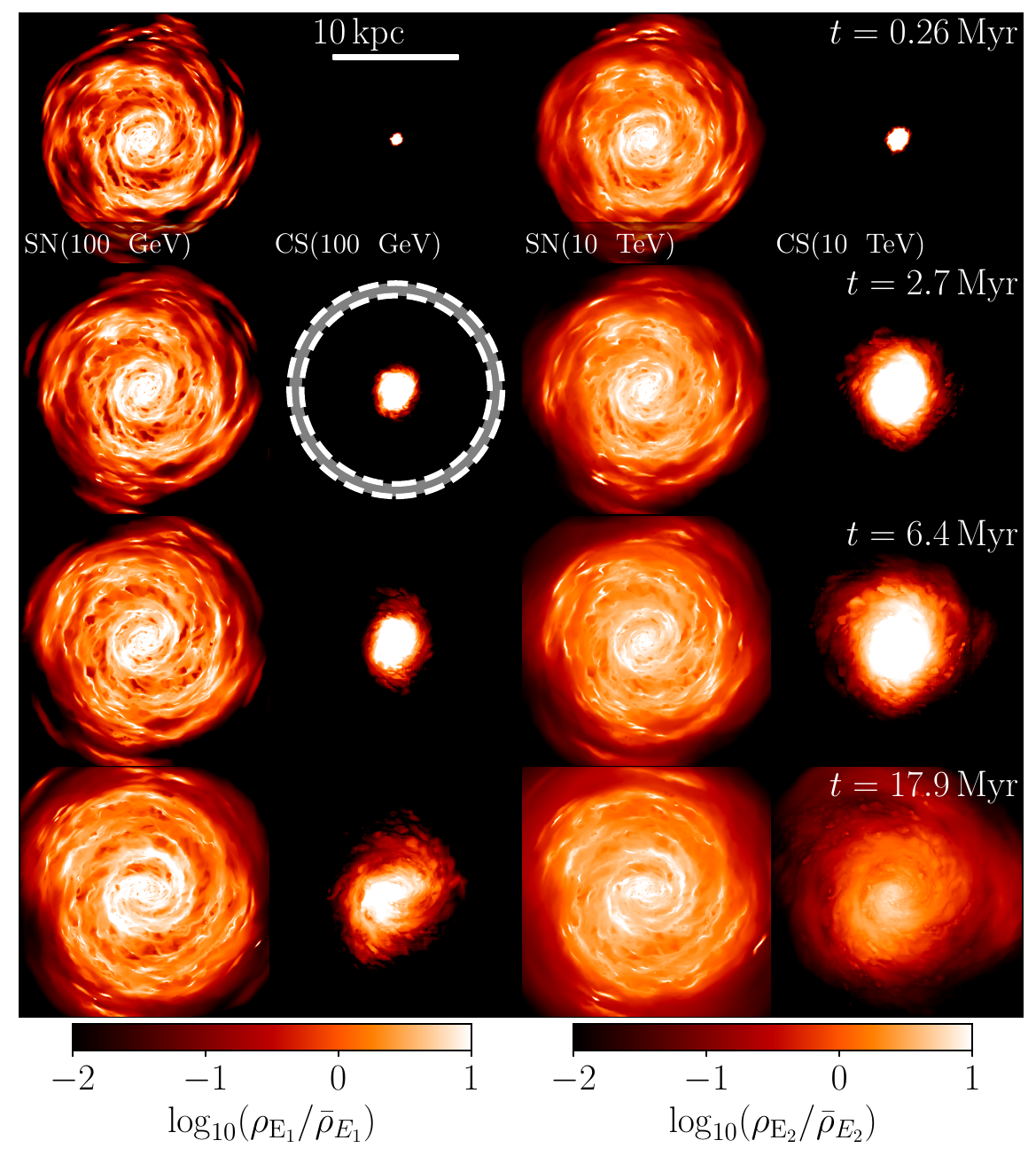}
    \caption{ Evolution of the several CR groups at four different times after injection of a pulse from the central source. The first and third columns present the CR groups energy density maps from SNe, at 100 GeV and 10 TeV respectively, while the second and fourth rows presents the same quantity for the CR groups associated with the central source. Each maps are normalized by the mean energy density in the SNe group at $t=0$. The grey annulus in the second column represents the solar neighborhood as defined previously, and spans between 7.5 and 8.5 kpc from the center of the MW. (\href{https://youtube.com/shorts/g6BHO2Rn_60}{Movie online})}
    \label{fig:sequence}
\end{figure*}

\subsection{Cosmic ray injection and diffusion}

To assess the impact of a galactic central source (CS) of CR on the overall CR spectrum measured at the solar neighborhood, we follow the propagation of two distinct populations of CR, injected by either the SNe or the CS, each of them being split into two energy groups.

In the following, we define the solar neighborhood as the ensemble of cells of the simulation whose position fulfill $-0.5 \text{ kpc} < z < 0.5 \text{ kpc}$ and $7.5 \text{ kpc} < r < 8.5 \text{ kpc}$, with $z$ is the axis orthogonal to the galactic disk and $r$ is the radial position of the cell.

\subsubsection{Energy groups and injection spectrum}

We follow two energy groups of CR, defined as intervals of kinetic energies following:

\begin{equation}
    \begin{aligned}
    &10^{1.5} \leq E_{\rm{CR}, 1} \leq 10^{2.5} {\rm{ \  GeV}} \\ \quad &10^{3.5} \leq E_{\rm{CR}, 2} \leq 10^{4.5} {\rm{ \ GeV}}.
    \end{aligned}
\end{equation}

\noindent These groups will be later on referred either as $E_1$ or $E_2$, or as their central value (100 GeV or 10 TeV respectively). We assume that the injection spectrum from sources, namely the SN and the CS, follow a power-law given by:

\begin{equation}
    \label{powerlaw}
    \frac{\text{d} n(K_{\rm{CR}})}{\text{d}K_{\rm{CR}}} \propto K_{\rm{CR}}^{-\alpha},
\end{equation}

where $n(K_{\rm{CR}})$ is the volumetric density of CR, $K_{\rm{CR}}$ is the kinetic energy of the CR, and $\alpha$ is the intrinsic source injection spectrum.

\noindent For each SN event, the amount of CR energy theoretically injected into the whole CR energy spectrum is $\eta_{\rm{SN}} \times e_{\rm{SN}}$, with $\eta_{\rm{SN}} = 0.1$ \citep{Dermer_2013}. The fractions of this energy to be injected in our two energy bins $E_1$ and $E_2$ is then calculated from Eq. (\ref{powerlaw}), taking a slope of $\alpha = 2.4$ \citep{Lagutin_2014,Blasi_2012}. \\

\noindent We follow the same procedure for the central source. A fraction $f_{\rm{CS}}=0.1$ of $E_F = 10^{56}$ erg, the minimum total energy required to form the FB is assumed to be redistributed as CR energy. Note that this amount of injected CR energy can be re-scaled in post-processing to explore the effect of the exact value of $E_F$, see Sect. \ref{howmuch}. The corresponding amount of energy to be injected in each energy group is obtained from the power-law assuming $\alpha = 2.4$, accordingly with measurements assuming hadronic model for the gamma-ray emission of the FB \citep{Ackermann_2014}. Several energy pulses can be injected with a fixed time interval $\Delta t$.

\subsubsection{Energy losses}

In this study, we only evaluate the propagation of the CR in the Galactic disk, and not its effect on the ISM. Thus we intentionally neglect any feedback process on the gas and possible energy losses. In particular, we do not consider the cooling of proton CR through pion production from proton-proton collision. The typical cooling timescale for this process is given by:

\begin{equation}
    t_{\rm{pp}} = \frac{1}{\kappa_{\pi^0}n_{\rm{H}}\sigma_{\rm{pp}}c} \simeq 5 \cdot 10^{15} \times \left( \frac{n_{\rm{H}}}{1 \, \rm{cm^{-3}}} \right)^{-1} \, \rm{s},
\end{equation}

where $\kappa_{\pi^0} \simeq 0.2$ is the fraction of the protons energy transfered to the pions, $\sigma_{\rm{pp}} \simeq 40 \, \rm{mb}$ is the collision cross section and $n_{\rm{H}}$ is the hydrogen number density \citep{Rieger}. Taking $n_{\rm{H}} = 10 \, \rm{cm}^{-3}$ as an upper limit for the gas density in our simulated MW, we obtain a typical cooling timescale of $t_{\rm{pp}} \sim 10^{7}$ yr, which is also the typical duration of our simulations. The results presented in this paper should thus be considered as an upper limit for the contribution of the CS to the local CR population.

\subsection{Simulations parameters}

In total, we run 6 simulations whose various parameters are summarized in Table \ref{table:simuparam}. The diffusion coefficient of the CR in the ISM for each simulation are fixed and taken assuming a power law of slope $\beta=0.3$ \citep{Strong_2007}:

\begin{equation}
    D(E) = 10^{28} \, \text{cm}^2 \, \text{s}^{-1} \left( \frac{E}{1 \, \text{GeV}} \right)^\beta.
\end{equation}

The obtained values for the 100 GeV and 10 TeV energy groups are $10^{29}\, \text{cm}^2 \, \text{s}^{-1} $ and $7 \times 10^{29}\, \text{cm}^2 \, \text{s}^{-1} $, respectively. Since solving the anisotropic diffusion equation of the CR is particularly computationally expensive, our 5 main simulations are run with a maximum resolution of 29 pc. Two additional simulations with a maximum resolution of 15 pc and 59 pc were also run to evaluate numerical convergence. Finally, we perform one additional run with isotropic CR diffusion for comparison.

\begin{table*}
\centering
\caption{Parameters of our simulations.}
\begin{tabular}{lcllc}
\hline\hline
\multicolumn{1}{c}{Name} & Res.  & Propagation Type & Injection & $\Delta t_p$ \\ \hline
s15pc                    & 15 pc & anisotropic      & single pulse        & -            \\
s29pc (\textit{fiducial})         & 29 pc & anisotropic      & single pulse        & -            \\
s29pc-iso                & 29 pc & isotropic        & single pulse        & -            \\
m29pc-1myr               & 29 pc & anisotropic      & multiple pulses        & 1 Myr        \\
m29pc-3myr               & 29 pc & anisotropic      & multiple pulses        & 3 Myr        \\
s59pc                    & 59 pc & anisotropic      & single pulse        & -  \\ 
\hline
\end{tabular}
\label{table:simuparam}
\end{table*}

\section{Results}

\subsection{Disk propagation and outflows}

Projections of the CR energy density in both SN and CS populations for our s15pc run and at four different times are presented in Fig. \ref{fig:sequence}. Variable $t$ here refers to the time spanned since the injection of the unique pulse of $10^{57}$ erg. The 100 GeV CR are found to diffuse much slower than the 10 TeV group, which is expected due to their lower diffusion coefficient. After $\sim$ 18 Myr, most 10 TeV CR have diffused out the Galactic center, while a large fraction of the 100 GeV CR is still confined within the bar.

\begin{figure}[h!]
    \centering
    \resizebox{0.95\hsize}{!}{\includegraphics{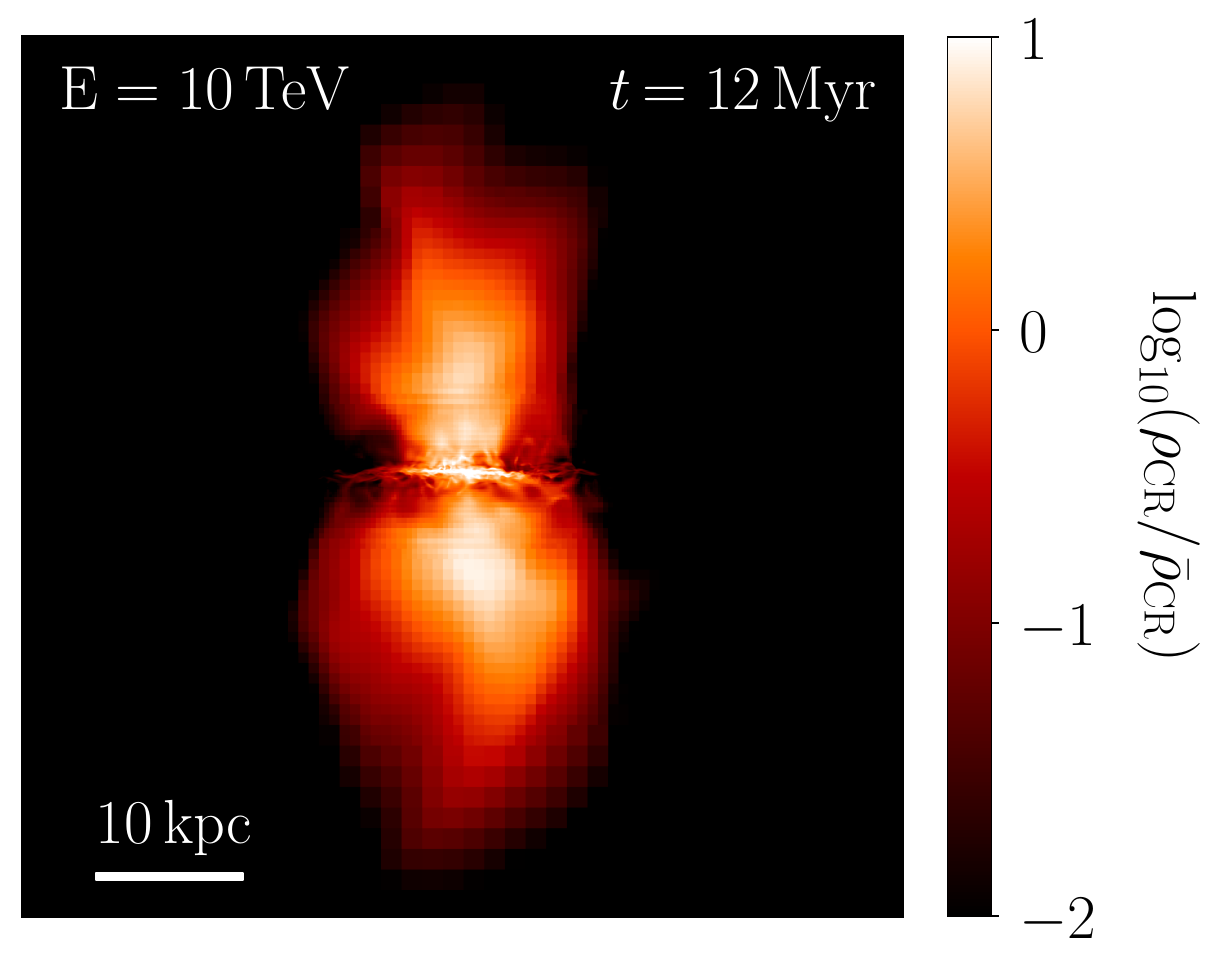}}
    \caption{Edge-on view of the 10 TeV CR group energy density injected by the CS in our s15pc run, 12 Myr after injection. As visible, bubble-like structures naturally emerge is our model, due to the topology of the magnetic field.}
    \label{fig:bubble}
\end{figure}

Since we want to quantify the effect of CR within the disk, we first evaluate how much of the injected energy is staying within the disk and conversely what fraction of the CR energy is outflowing from the Galactic disk. As visible in Fig. \ref{fig:bubble}, significant amount of CR energy is propagating out of the disk, and forms large bubble-shaped structures in both sides of the galactic plan. Remarkably, the spatial extent of these outflowing structures is consistent with estimated size of the FB \citep{Su_2010}, which we evaluate to roughly 10 kpc at $t=3$ Myr.

To evaluate the impact of the outflow on the overall CR energy budget in the disk, we introduce $f_{i,\text{outflow}}$, the fraction of the energy injected by the CS escaping the disk for the energy group $E_i$, defined by:

$$f_{i,\text{outflow}}(t) = 1 - \frac{\sum_{j \in \text{disk}} \, \rho_{i,j}(t) \cdot V_j}{\sum_{j \in \text{box}} \, \rho_{i,j}(t) \cdot V_j},$$

with $\rho_{i,j}(t)$ the CR energy density in cell $j$, for the $E_i$ energy group and from the CS population. $V_j$ designates the volume of the cell $j$. Here, the disk is defined as the set of gas cells contained in a slice of thickness 1 kpc vertically centered on the midplane of the box.

\begin{figure*}[h!]
    \centering
    \includegraphics[width=\textwidth]{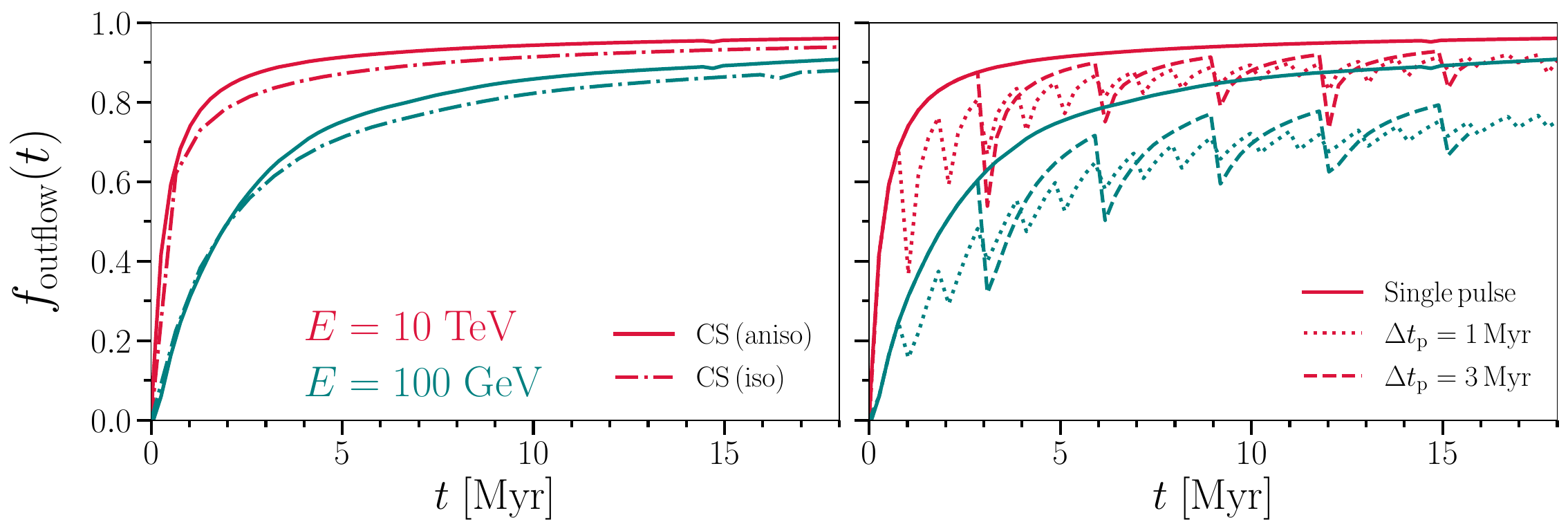}
    \caption{Evolution of $f_{\text{outflow}}$ as a function of time and for various parameters. On the left panel we present the influence of the propagation model (i.e. either anisotropic or isotropic) for our 29 pc resolution, single pulse run (s29pc). On the right panel we show the influence of the intermittence of the source.}
    \label{fig:outflow_fraction}
\end{figure*}

Results are shown in Fig. \ref{fig:outflow_fraction}. The choice of the propagation mode is found to have minor influence on the fraction of energy staying in the disk as more than 80 \% of the injected CR energy escape from the disk after at most 8 (resp. 2.5) Myr for the $E_1$ ($E_2$) energy bin for both anisotropic and isotropic runs. Intermittence is found to increase the proportion of energy contained in the disk, as recurrent injection from the CS constantly refuel the disk with additional energy that did not yet had time to outflow from the disk. One can verify this assumption by computing the diffusion length $L_\text{diff}$ after 1 Myr, the time between two consecutive pulses:

$$L_\text{diff} = \sqrt{\Delta t_p \cdot D_\parallel} \sim 0.5 \text{ kpc},$$

which is of the same order as the half-height of the galactic disk. In any case, the CR energy escaping the disk and filling the bubble-like structures shown in Fig. \ref{fig:bubble} is expected to largely dominate the total CR energy injected by the CS. We also observe that the magnetic field topology might locally redirect outflowing streams of CR back into the disk. However, due to the Eulerian nature of {\scshape{Ramses}}, this effect remains hard to quantify.

\subsection{How much does the central source contribute to the local cosmic-ray population ?}
\label{howmuch}

\begin{figure}[h!]
    \centering
    \resizebox{\hsize}{!}{\includegraphics{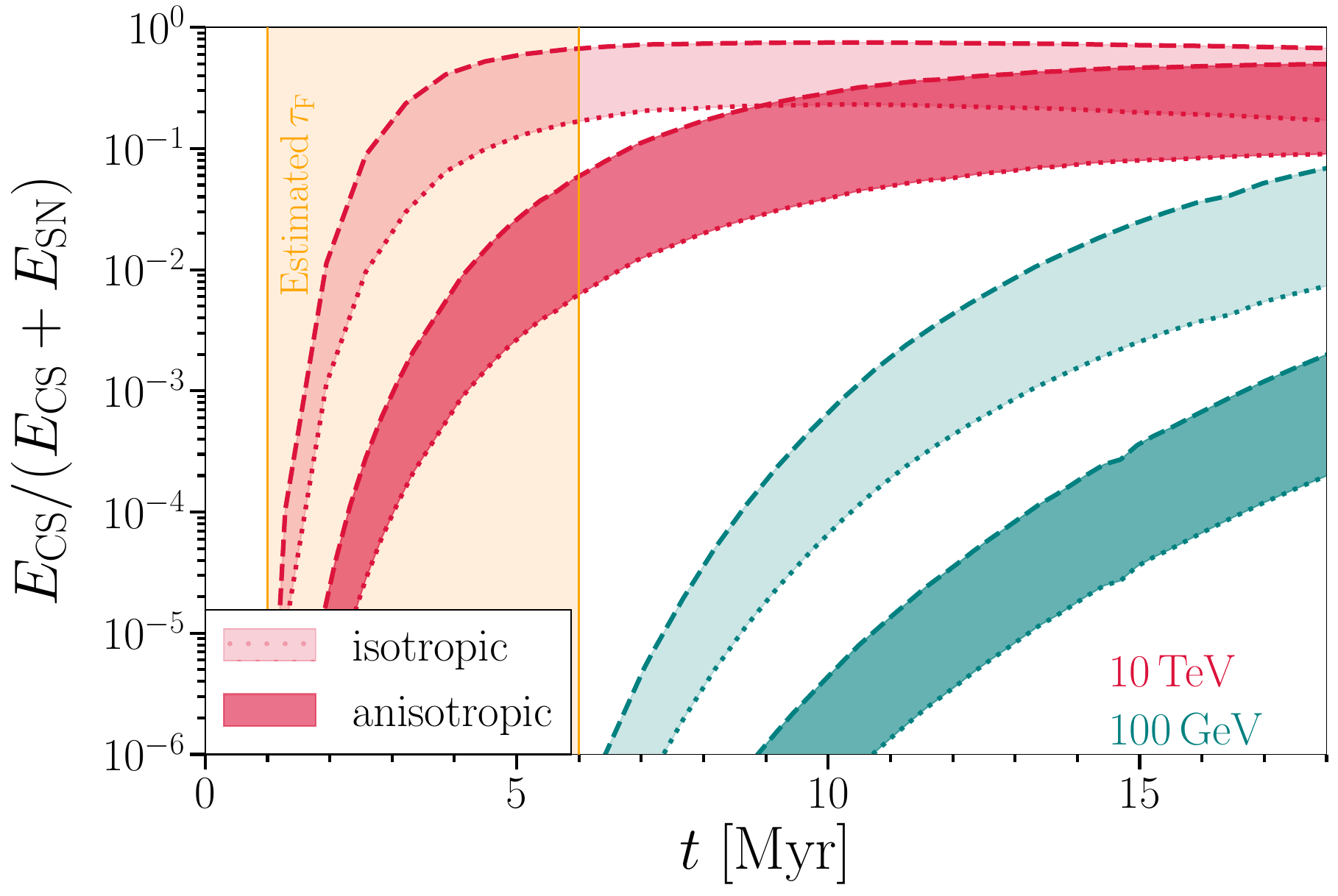}}
    \caption{Proportion of the total CR energy measured at the solar neighborhood attributed to the CS source, as a function of time and for each energy bin. Since current estimates of the energy required to form the FB spans between $10^{56}$ and $10^{57}$ erg, this quantity is left as a free parameter, the obtained range of values being represented by the colored surfaces. For each of them, the dotted (dashed) line represents the scenario in which $10^{56}$ erg ($10^{57}$ erg) is required to form the FB. The estimated age of the FB (1--6 Myr) \citep{Guo_2012,Zhang_2020} is represented by the orange surface. }
    \label{fig:constraint}
\end{figure}

In Fig. \ref{fig:constraint}, we evaluate the relative contribution of a single CS pulse in the total amount of CR energy measured in the solar neighborhood. Because the total energy $E_F$ required to form the FB is not well constrained and is thought to range from $10^{56}$ to $10^{57}$ erg, we re-scale the amount of energy injected from the CS in post-processing to explore this whole interval. The resulting range of values is represented as the colored surfaces. The dotted (dashed) line corresponds to the case where $E_F = 10^{56}$ erg ($10^{57}$ erg) is injected from the CS. The orange surface shows the estimated age $\tau_F$ of the FB, which spans from 1 to 6 Myr \citep{Guo_2012,Zhang_2020}. For each of the two energy bins, the values obtained in the case where the CR propagate isotropically is shown by the shaded surface.

In the lower energy bin, the contribution of the CS is strictly negligible at any time in the $\tau_F$ estimate range. This contribution might eventually increase up to a few percent in the isotropic case after 15 Myr, much beyond the larger estimate of $\tau_F$. For the higher energy bin, the contribution of the CS reaches up to a few percents within the first 6 Myr in the anisotropic case, and up to 70 \% in the isotropic case. We conclude that the current estimates of the age of the FB and of the total CR energy injected from the CS implies that the CS is not able contribute significantly to the observed total CR energy budget. We note that this results from the anisotropic CR diffusion, and that considering an isotropic diffusion significantly change the result. Taking the source intermittence into account do not change these conclusions.

We emphasize that our simulation with isotropic diffusion provides results consistent with previous theoretical work \citep{Jaupart_2018}, which are thus likely to overestimate the CS contribution at the solar neighborhood by up to several orders of magnitude. 

\subsection{Impact on the slope of the spectrum}

\begin{figure}[h!]
    \centering
    \resizebox{\hsize}{!}{\includegraphics{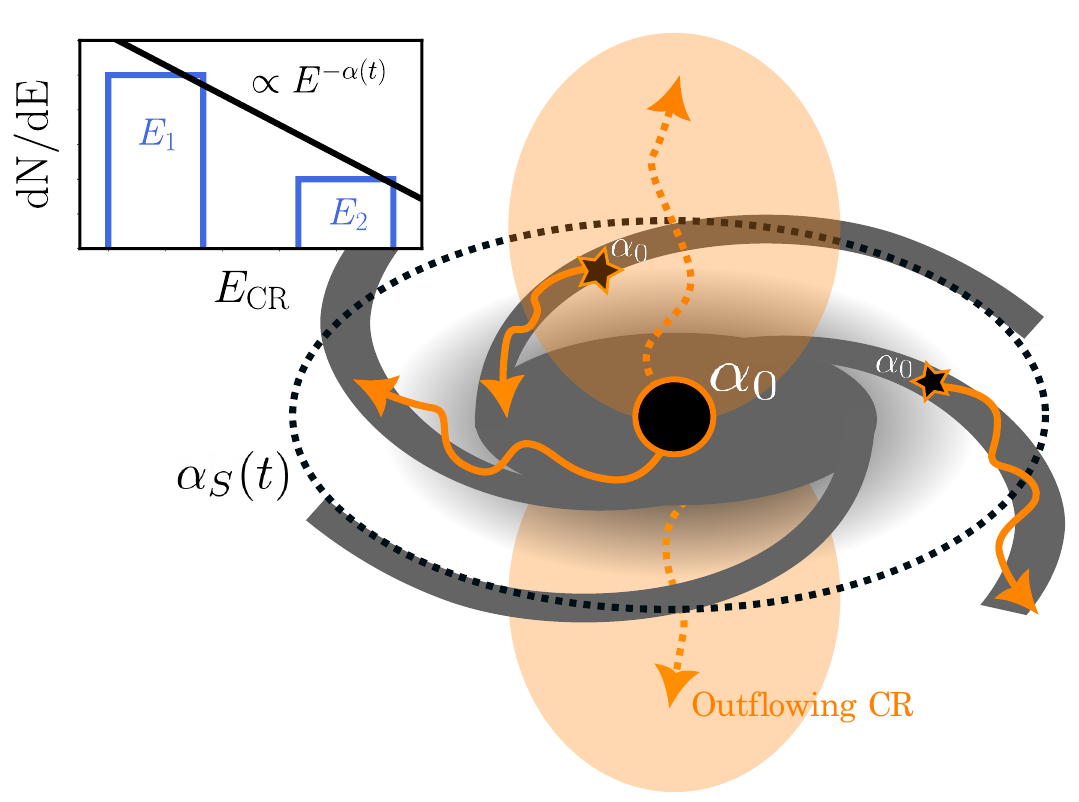}}
    \caption{Diagram illustrating the principle of measurements of the spectrum slope. The central SMBH ($\bullet$) particle and the SNe ($\star$) inject CR with an initial spectrum of slope $\alpha_0$. While propagating along the magnetic field lines, the slope of this spectrum can exhibit some modifications due to the energy dependence of the diffusion coefficient, and to a fraction of the CR escaping the disk. The spectrum's slope at solar neighborhood $\alpha_S(t)$ (later simply denoted as $\alpha$) is then measured as a function of time for both the SNe and the CS populations by comparing the relative amount of energy contained in each energy bin, averaged over all solar neighborhood gas cells.}
    \label{fig:schema}
\end{figure}

To assess the effect of the central source on the CR flux in the solar neighborhood, we compute the slope of the spectrum using the energy density populating our two energy bins within all the cells contained in the solar neighborhood. The configuration is illustrated in Fig. \ref{fig:schema}.

The slope $\alpha$ of the CR spectrum in the solar neighborhood ($\odot$) is then obtained following:

\begin{equation}
    \alpha(t) = \frac{1}{\log E_2 / E_1} \log \left( \frac{\sum_{j \in \odot} \, \rho_{E_1,j}(t) \cdot V_j }{ \sum_{j \in \odot} \, \rho_{E_2,j}(t) \cdot V_j} \right).
\end{equation}

\begin{figure*}[h!]
    \centering
    \includegraphics[width=\textwidth]{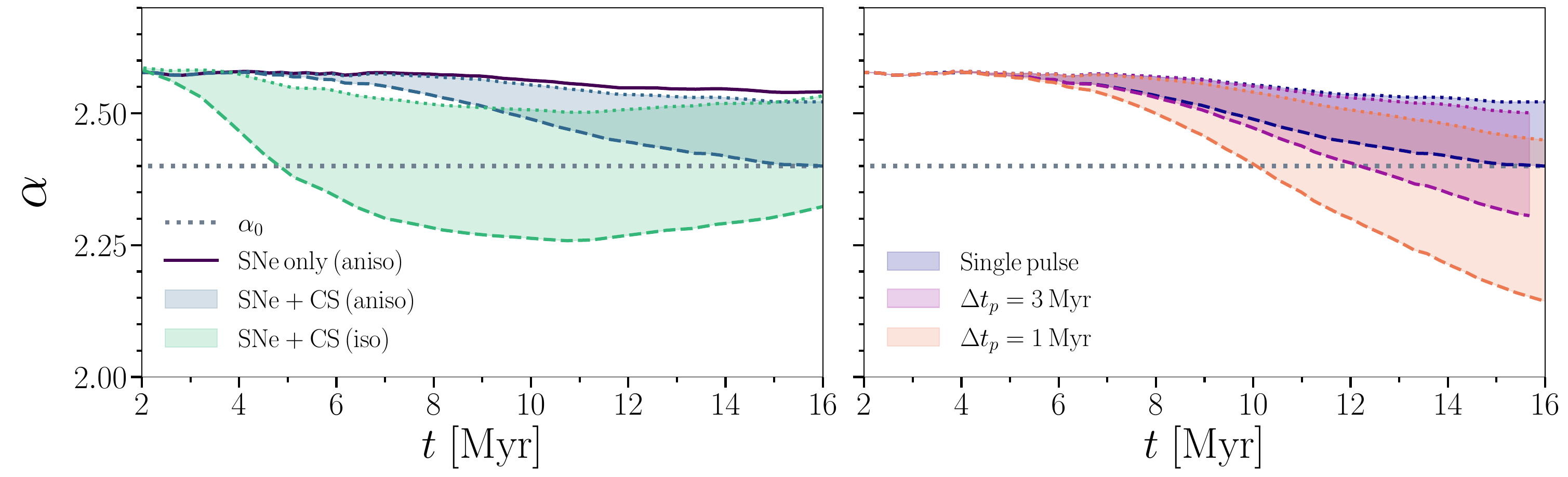}
    \caption{Evolution of the spectrum slope under various conditions. Because the exact quantity of energy injected at each pulse is only weakly constrained, this quantity is left as a free parameter, which can vary between $10^{56}$ and $10^{57}$ erg. The corresponding range of values obtained are representend by the colored surfaces, while the dotted (dashed) line represents the case where a total CR energy of $10^{56}$ erg ($10^{57}$ erg) is injected from the CS at $t=0$. The left panel shows the influence of the propagation mode (i.e. isotropic or anisotropic). The right panel shows the influence of the source intermittence. The results of two simulations with intermittent source are presented, with injection periods of 1 Myr and 3 Myr. The anisotropic, single injections is also represented by the solid line for comparison.}
    \label{fig:slope}
\end{figure*}

Here, the logarithmic term is the ratio of the CR energy contribution from both SN and the CS contained in the solar neighborhood for the two energy bins. It is obtained by summing the CR energy contained in each cell of the solar neighborhood, here labelled with index $j$. The evolution of the slope is presented in Fig. \ref{fig:slope}. Here, the exact quantity of energy injected by the central source is left as a free parameter, which can vary between $10^{56}$ and $10^{57}$ erg. The colored surfaces represents the range of values covering this whole interval, while the colored lines represents the median values. The left panel shows the influence of the propagation mode on the evolution of the slope for the emission of a single pulse at $t_0$. The grey dotted line presents the injection spectrum $\alpha_0 = 2.4$. The purple curve present the slope of the CR spectrum when only accounting for the two SNe groups. It is rather constant with a mean value of $\Bar{\alpha}=2.6$, which is consistent with measured values of the average spectrum in the local Galaxy, i.e. in the ISM within 1 kpc distance from the Sun \citep{Neronov_2017}. The green line shows the slope of the total CR spectrum, i.e. containing both the contribution of the SNe and the CS, for the isotropic propagation mode, while the blue line indicate the same quantity for the anisotropic propagation mode. 

The propagation mode is found to have a critical impact on the ulterior shape of the spectrum. In the isotropic case, the spectrum becomes significantly harder after $\sim 10$ Myr. This hardening effect is due to the fact that the CR belonging to the higher energy group $E_2$ diffuse faster than the $E_1$ group due to their larger diffusion coefficient. They thus populate quicker the solar neighborhood, leading to a flattening of the spectrum. In the anisotropic case, the propagation is more sensitive to the complex morphology of the Galaxy, since the CR propagate preferentially along magnetic field lines which follows its spiral shape (see Fig. \ref{fig:condini}). Direct effect observed here is weaker impact on the spectrum's slope, which is likely to be undetectable for the first 6 -- 8 Myr after the energy injection.

We also model the eventuality that the central source of CR is intermittent, and present the result of our two associated simulations, with injection periods of 1 Myr and 3 Myr. Results are presented in the right panel of Fig. \ref{fig:slope}. Remarkably, the latency period between the energy injection and the time at which the spectrum at solar neighborhood starts being modified by the contribution of the central source CR is found to be independent from the source period, and roughly equal to 6 Myr. At later times, the effect on the slope is increasing with the frequency of injection from the central source. 15 Myr after the first injection, modification of the slope is found to span from $\sim 7 \%$ for the single pulse simulation, to 21 $\%$ for the intermittent run with $\Delta t_p = 1$ Myr. Considering that current measurements of the CR spectrum slope have a precision of a few percents \citep{Neronov_2017}, the contribution of the central source in run with intermittent injection is unlikely to be measurable less than 10 Myr after the first injection. Since most estimations of the FB age ranges between 1 -- 6 Myr, we can though conclude that a significant contribution of CR emitted from the Galactic center to the overall CR spectrum measured in the solar neighborhood is unlikely to be observed by current surveys. These conclusions could however be modified if further observational work manage to identify evidence of older outburst, i.e. with estimated ages superior to 10 Myr.

\section{Conclusion}

We have investigated whether CR outbursts from the Galactic center are likely to contribute significantly to the CR spectrum observed at the solar neighborhood. By comparing simulations run with isotropic and anisotropic diffusion, we conclude that the diffusion mode strongly influences how much the CS will contribute to the local CR flux. When taking into account the preferential diffusion of CR along magnetic field lines, CR are found to require a much longer timescale to reach the solar neighborhood. By comparing the typical timescale required for the CR to significantly impact the local CR flux slope with the estimated age of the FB in the jet-driven scenario, we conclude that the period of strong activity from the Galactic center suspected to explain the observations made by Fermi and eRosita telescopes is unlikely to affect the currently measured CR flux up to the TeV regime. Nonetheless, these conclusions do not exclude the possibility that hypothetical active phases of the Galactic center at older times might have a contribution to the local CR spectrum.

We emphasize that our CS injection scheme, as well as our propagation model remain simplified. We do not include some internal properties of the source, such as its possible anisotropy, stochasticity nor continuity. We also do not include production and acceleration of CR within and at the edge of the outflowing bubbles, as well as CR re-acceleration, streaming, energy loss or secondary CR production. Further research work should focus on evaluating whether acceleration of CR within the shock waves observed by eRosita is likely to challenge our conclusions. Treatment of CR including a larger number of energy bins and energy losses might also help constraining more precisely the possible features of such event on the local CR flux. Finally, more observational constraints might help improving the modeling of the CS, especially regarding its intrinsic injection spectrum.

\begin{acknowledgements}
We thank Alexandre Marcowith, Joakim Rosdahl, Yohan Dubois, Etienne Jaupart, Volker Heesen, Marcus Brüggen, Franco Vazza and Maria Werhahn for insightful discussions. We thank Eric Emsellem who provided us with the initial conditions for the Milky Way. This work was performed thanks to HPC resources provided by the PSMN (Pôle Scientifique de Modélisation Numérique) of the ENS de Lyon. We also thank Florent Renaud for providing the open-source visualization tool {\scshape{rdramses}}.
\end{acknowledgements}

%
%

\bibliographystyle{aa} 
\bibliography{biblio} 

\clearpage

\begin{appendix}
\section{Energy conservation and numerical convergence}

\begin{figure}[h!]
    \centering
    \resizebox{\hsize}{!}{\includegraphics{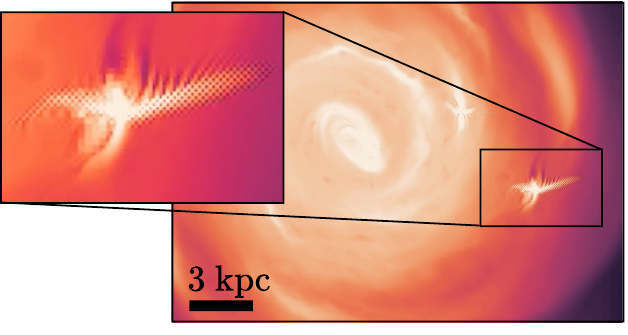}}
    \caption{CR energy density projection for one of our preliminary, low-resolution run. Several artifact of diffusion produced by injection from SNe are visible. Typical feature of such event is a spike of CR energy propagating perpendicularly to the magnetic field lines, and endowed with chessboard pattern.}
    \label{fig:chessboard}
\end{figure}

While running our first simulations, we found non-physical production and destruction of CR energy in our simulated box, which we associated with wrong estimates of CR fluxes at cells interfaces. Visual artifacts known as chessboard patterns are usually visible under these conditions and are shown in Fig. \ref{fig:chessboard}. The origin of these patterns is discussed in \cite{Sharma_2007}.

\begin{figure}[h!]
    \centering
    \resizebox{0.85\hsize}{!}{\includegraphics{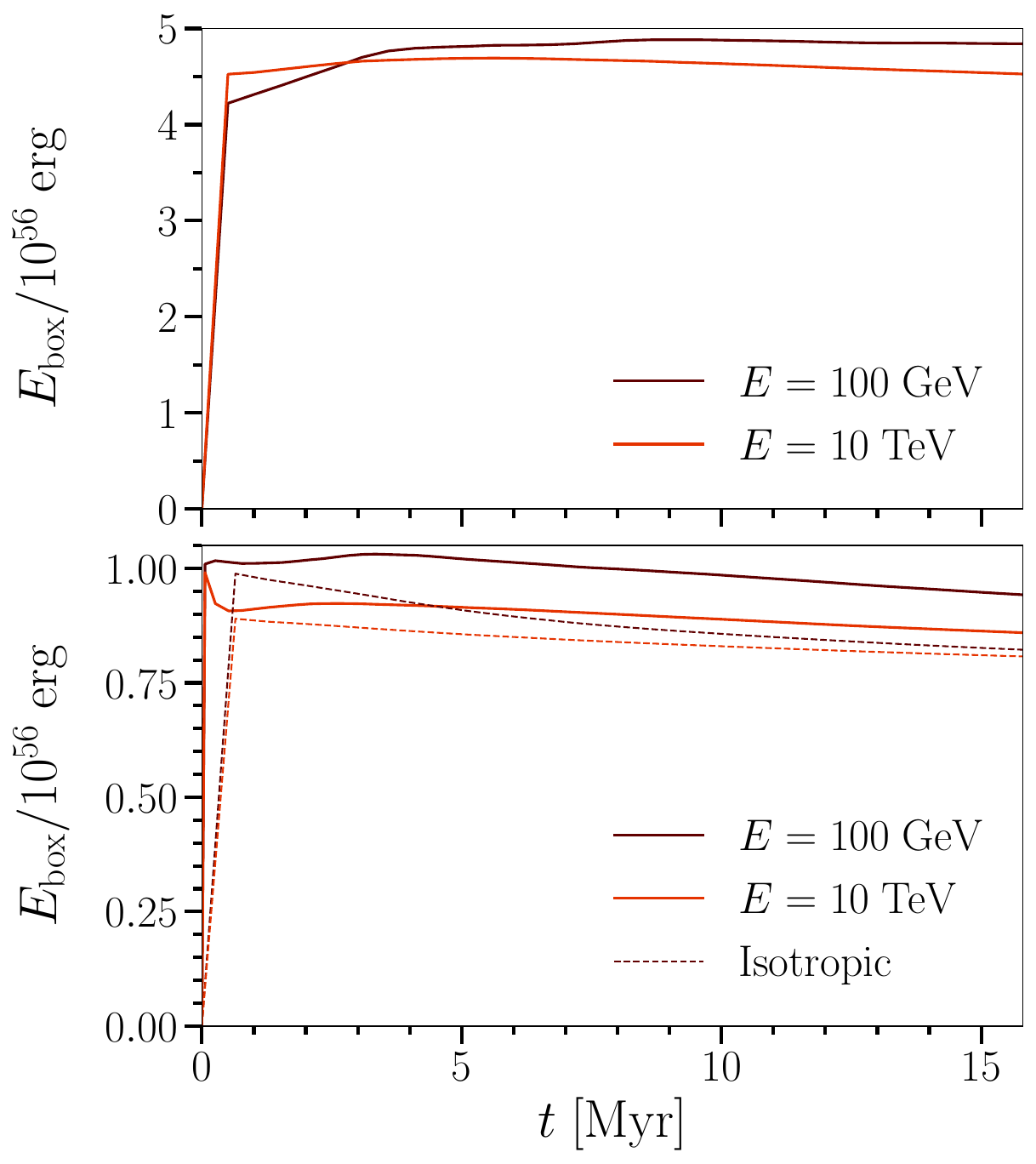}}
    \caption{Comparison of energy conservation between test runs without (top panel) and with (bottom panel) flux limiter. Test runs are performed by injecting a pulse of energy $10^{56}$ erg in both energy bins at $t = 0$ Myr. The total CR energy contained in the simulated box is measured at various times for each of the two energy bins.}
    \label{fig:energy_conservation}
\end{figure}

These sources of non-conservation, known to dramatically impact the reproducibility of our results, can be efficiently reduced by the use of flux limiters. Energy conservation with and without the use of slope limiter is presented in Fig. \ref{fig:energy_conservation}. For these tests, we theoretically inject a pulse of $10^{56}$ erg in both energy bins. Maroon and orange curves represents the energy effectively injected in the 100 GeV and the 10 TeV groups, respectively. As visible on the left panel, numerical artifacts lead to dramatic energy production, which multiple the amount of energy theoretically injected by a factor of nearly five. After activating the slope limiter, we obtain results presented on the right panel, which fulfill energy conservation over our tests run by nearly 90 \% in both energy groups for the anisotropic runs. We verified that the subsequent energy loss, likely associated to non-conservation due to Dirichlet boundary conditions at coarse-to-fine level interface \citep{Commercon_2014} only weakly affect our results. 

\begin{figure}[h!]
    \centering
    \resizebox{0.83\hsize}{!}{\includegraphics{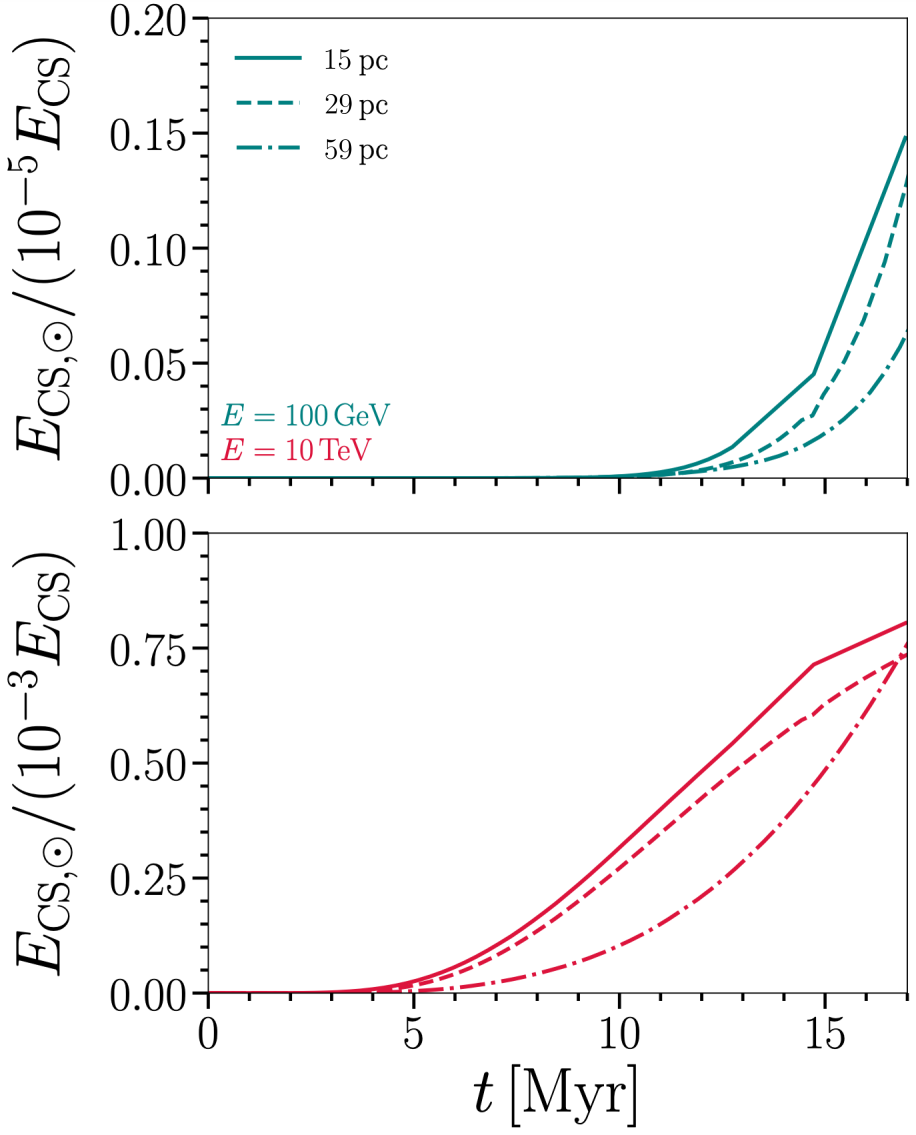}}
    \caption{Numerical convergence test. A single pulse of $E_{\text{CS}} = 10^{56}$ erg is injected from the central source and the total amount of CR energy contained in the solar neighborhood cells is followed through time. }
    \label{fig:convergence}
\end{figure}

Numerical convergence is verified by running several simulations with the same physical parameters and different numerical resolutions, namely 59 pc, 29 pc and 15 pc. In Fig. \ref{fig:convergence}, we present the total amount of energy measured in the solar neighborhood for our three single pulse simulations, namely with 59 pc, 29 pc and 15 pc spatial resolution. While not perfectly matching, 29 pc and 15 pc seems to come to a reasonable agreement. For this reason, we derive most of our results in this paper from our 29 pc resolution runs, which allowed us to run more simulations in the limited time we had for this project.

We emphasize that physical properties of our simulated Galaxy cannot be identical from one run to an other. The stochasticity of star formation, as well as the resolution-dependent effect of SNe on the gas dynamics leads to star formation rates which can differ by up to $\pm$ 0.5 M$_\odot$ yr$^{-1}$ from one resolution to an other. Subsequently, the morphology of the Galaxy, as well as the topology of its magnetic field vary from one run to an other, and can affect CR kinematics. This variability is likely to justify the lack of perfect convergence between our 29 pc and 15 pc resolution runs. Also, the amount of energy measured in the solar neigborhood in the lowest energy bin (100 GeV) is only a very small fraction of the injected energy. Dependence to morphological properties of the simulated Galaxy and its magnetic field is thus likely to be much larger than for the highest energy bin.

\end{appendix}

\end{document}